\documentclass[amsmath,amssymb,aps,pra]{revtex4-2}
\usepackage{xcolor,graphicx}        
\usepackage{braket}
\usepackage{float}
\usepackage{tikz}
\usepackage{braket}
\usepackage{hyperref}
\hypersetup{
	colorlinks = true,
	linkcolor = cyan,
	anchorcolor = cyan,
	citecolor = cyan,
	filecolor = cyan,
	urlcolor = cyan}

\begin{document}
% Leave a blank line between paragraphs instead of using \\
\title{The anisotropic quantum Rabi model with diamagnetic term}

\author{Jorge A. Anaya-Contreras$^{2}$, Irán Ramos-Prieto$^{1}$, Arturo Z\'uñiga-Segundo$^{2}$,  and Héctor M. Moya Cessa$^{1,}$*}
\affiliation{$^{1}$Instituto Nacional de Astrofísica Óptica y Electrónica, Calle Luis Enrique Erro No. 1, Santa María Tonantzintla, Puebla, 72840, Mexico\\
$^{2}$Instituto Politécnico Nacional, ESFM, Departamento de Física. Edificio 9, Unidad Profesional “Adolfo López Mateos”, 07738 CDMX., Mexico }

\begin{abstract}

%%% Leave the Abstract empty if your article does not require one, please see the Summary Table for full details.

We employ a squeeze operator transformation approach to solve the anisotropic quantum Rabi model that includes a diamagnetic term. By carefully adjusting the amplitude of the diamagnetic term, we demonstrate that the anisotropic Rabi model with the $A^2$ term can be exactly reduced to either a Jaynes-Cummings or an anti-Jaynes-Cummings model without requiring any approximations. 
\end{abstract}
\maketitle
\section{Introduction}

The interaction of atoms with cavity fields \cite{JC_1963,Gerry_Book,Larson_2022} is of great importance not only due to the fundamental questions that may be answered, but also because of the possible technological applications \cite{Nilakanta,Nilakanta2,Nilakanta3} as it has  entanglement, at the core of such interaction, which is the key ingredient of quantum information processes.

When analyzing this interaction several approximations are done, namely, the diamagnetic term \cite{Crisp,Crisp_1993,Kockum_2019b,Salado_2021} is dropped, the dipole and rotating wave approximations are made and the interaction with environments \cite{MoyaCessaPRA} is not considered, this is, studies are focused on high-$Q$ cavities. However, there are intensity regimes where such approximations are not any more valid and then it is needed to consider the full interaction, {\it i.e.}, the quantum Rabi model~\cite{Braak}. Solutions for this problem have been already provided \cite{Swain,Moya_2000,Chen_2011,Braak}, usually in terms of infinite continued fractions \cite{Swain}.

It has been shown that the diamagnetic term  may be of importance in the deep-strong-coupling (DSC)  and ultra-strong-coupling regimes (USC) \cite{Kockum_2019b}. In the atom-field interaction, the diamagnetic term is usually dropped as it is a term that it is of the order of counter rotating terms \cite{Crisp_1993}. However, in other regimes the impact of the diamagnetic term is  non-negligible  and it may become dominant in the DSC regime \cite{Kockum_2019b}.

Generalizations of the quantum Rabi model, such as the anisotropic quantum Rabi model \cite{Xie,Ye,Boutaka}, have been studied. In particular it has been shown the existence of entanglement \cite{Boutaka} and antibounching-to-bounching transitions of photons \cite{Ye}.

In this contribution we show that an anisotropic Rabi model that includes the diamagnetic term may be reduced, by using a transformation that involves the squeeze operator \cite{Satyanarayana_1989}, to the Jaynes-Cummings \cite{JC_1963} and anti-Jaynes-Cummings models \cite{Anti}. These kind of systems have been shown to have partner Hamiltonians in the theory of supersymmetry (SUSY) \cite{Ivan,David_14} that allows the connection of physical models via supersymmetric operators, {\it i.e.},  mapping the corresponding Hilbert spaces.

\section{The anisotropic quantum Rabi model}
The Hamiltonian for the anisotropic quantum Rabi model, including the diamagnetic term, can be expressed as (with $\hbar = 1$):
\begin{equation}\label{dia}
    \begin{split}
    \hat{H} &= \omega \hat{a}^\dagger \hat{a} + \frac{\omega_0}{2} \hat{\sigma}_z + \big(g_1 \hat{a} + g_2 \hat{a}^\dagger\big) \hat{\sigma}_+ + \big(g_1 \hat{a}^\dagger + g_2 \hat{a}\big) \hat{\sigma}_- + D \big(\hat{a} + \hat{a}^\dagger\big)^2, \\
    &= \omega_D \hat{a}^\dagger \hat{a} + \frac{\omega_0}{2} \hat{\sigma}_z + \big(g_1 \hat{a} + g_2 \hat{a}^\dagger\big) \hat{\sigma}_+ + \big(g_1 \hat{a}^\dagger + g_2 \hat{a}\big) \hat{\sigma}_- + D \big(\hat{a}^2 + \hat{a}^\dagger{}^2\big) + D,
    \end{split}
\end{equation}
where $\omega_D = \omega + 2D$. Here, $\hat{a}$ and $\hat{a}^\dagger$ represent the annihilation and creation operators of the bosonic field, satisfying the commutation relation $[\hat{a}, \hat{a}^\dagger] = 1$. The Pauli atomic operators $\hat{\sigma}_\pm$ and $\hat{\sigma}_z$ describe the two-level atomic system, obeying the commutation relations: $[\hat{\sigma}_+, \hat{\sigma}_-] = \hat{\sigma}_z$ and $[\hat{\sigma}_z, \hat{\sigma}_\pm] = \pm 2 \hat{\sigma}_\pm$. Here, $\omega$ and $\omega_0$ denote the field frequency and the atomic transition frequency, respectively, while $D$ quantifies the diamagnetic amplitude. The coupling constants $g_1$ and $g_2$ characterize the interaction strength between the atom and the field.

To eliminate the residual diamagnetic term $\left(\hat{a}^2 + \hat{a}^{\dagger2}\right)$ from Eq.~\eqref{dia}, we apply a unitary transformation defined by the squeeze operator:
$\hat{S}(r) = \exp\left[\frac{r}{2}\left(\hat{a}^2 - \hat{a}^{\dagger2}\right)\right]$~\cite{Gerry_Book}, where $r$ is the squeezing parameter to be determined subsequently. Under this transformation, the annihilation and creation operators transform as:
\begin{equation}
    \hat{S}^\dagger(r) \hat{a} \hat{S}(r) = \mu \hat{a} - \nu \hat{a}^\dagger, \quad
    \hat{S}^\dagger(r) \hat{a}^\dagger \hat{S}(r) = \mu \hat{a}^\dagger - \nu \hat{a}, \quad
    \text{where} \quad \mu = \cosh(r), \quad \nu = \sinh(r).
\end{equation}
Applying the transformation $\hat{H}_S = \hat{S}^\dagger(r) \hat{H} \hat{S}(r)$, the Hamiltonian becomes:
\begin{equation}
    \begin{split}
    \hat{H}_S &= \big[\omega_D \big(\mu^2 + \nu^2\big) - 4D \mu \nu\big] \hat{a}^\dagger \hat{a} + \frac{\omega_0}{2} \hat{\sigma}_z\\
    &\quad + \big(\mu g_1 - \nu g_2\big) \big(\hat{a} \hat{\sigma}_+ + \hat{a}^\dagger \hat{\sigma}_-\big) + \big(\mu g_2 - \nu g_1\big) \big(\hat{a} \hat{\sigma}_- + \hat{a}^\dagger \hat{\sigma}_+\big) \\
    &\quad + \big[D \big(\mu^2 + \nu^2\big) - \mu \nu \omega_D\big] \big(\hat{a}^2 + \hat{a}^\dagger{}^2\big)+\nu^2\omega_D-2D\mu\nu+D.
    \end{split}
\end{equation}
By imposing the condition $\frac{D}{\omega_D} = \frac{\mu \nu}{\mu^2 + \nu^2}$, the Hamiltonian simplifies to:
\begin{equation}\label{H_S}
    \begin{split}
    \hat{H}_S =&\, \omega_D\frac{\big(\mu^2 - \nu^2\big)^2}{\mu^2 + \nu^2} \hat{a}^\dagger \hat{a} + \frac{\omega_0}{2} \hat{\sigma}_z + \big(\mu g_1 - \nu g_2\big) \big(\hat{a} \hat{\sigma}_+ + \hat{a}^\dagger \hat{\sigma}_-\big) + \big(\mu g_2 - \nu g_1\big) \big(\hat{a} \hat{\sigma}_- + \hat{a}^\dagger \hat{\sigma}_+\big)\\
    &+\omega_D\left(\nu^2-\frac{2\mu^2\nu^2+\mu\nu}{\mu^2+\nu^2}\right).
    \end{split}
\end{equation}
Thus, the squeezing transformation eliminates the residual diamagnetic term, thereby simplifying the system to the anisotropic quantum Rabi model. In the special case where $g_1 = g_2$, the model reduces to the standard quantum Rabi model \cite{Braak}. However, this work focuses on the scenario where $g_1 \neq g_2$. Specifically, we investigate two distinct cases: (a) $g_2 < g_1$, which corresponds to the Jaynes-Cummings model, and (b) $g_1 < g_2$, associated with the anti-Jaynes-Cummings model~\cite{Anti,Ivan}. The Hamiltonian described by Eq.~(\ref{H_S}) represents one of the key contributions of this work, providing a comprehensive framework for exploring the interplay between anisotropy, squeezing, and light-matter interactions within the anisotropic quantum Rabi model.

%------------------------------------
\subsection{Jaynes-Cummings Model}
Once the Hamiltonian, Eq.~(\ref{H_S}), is established, we proceed to fix the parameters to recover the well-known Jaynes-Cummings model. This requires imposing the condition $\frac{\mu}{\nu} = \frac{g_1}{g_2}$, where $\mu = \cosh(r)$ and $\nu = \sinh(r)$ are the hyperbolic functions associated with the squeezing parameter $r$, and the parameter region determined by the condition $g_2 < g_1$. Under this condition, the Hamiltonian $\hat{H}_S$ simplifies to the Jaynes-Cummings Hamiltonian, which takes the form: 
\begin{equation}
    \hat{H}_{\text{JCM}} = \omega_{\text{eff}} \hat{a}^\dagger \hat{a} + \frac{\omega_{0}}{2} \hat{\sigma}_z + g_{\text{eff}} \left( \hat{a} \hat{\sigma}_+ + \hat{a}^\dagger \hat{\sigma}_- \right) + f_0,
\end{equation}
where the effective frequency $\omega_{\text{eff}}$, the effective coupling constant $g_{\text{eff}}$, and the zero-point energy shift $f_0$ are explicitly defined as:
\begin{equation}
  \omega_{\text{eff}} = \omega_D \left( \frac{g_1^2 - g_2^2}{g_1^2 + g_2^2} \right), \quad g_{\text{eff}} = \sqrt{g_1^2 - g_2^2}, \quad \text{and} \quad f_0 = \omega_D \left( \frac{g_1 g_2 - g_2^2}{g_1^2 + g_2^2} \right).
\end{equation}
For the specific case where $g_2 < g_1$, the condition eliminates the residual diamagnetic term, which takes the following form in this parameter regime:
\begin{equation}\label{condicion}
    \frac{D}{\omega_D} = \frac{\mu \nu}{\mu^2 + \nu^2} = \frac{\tanh(2r)}{2} = \frac{g_1 g_2}{g_1^2 + g_2^2}.  
\end{equation}
This expression highlights the relationship between the squeezing parameter $r$ and the coupling constants $g_1$ and $g_2$. Furthermore, the hyperbolic functions $\cosh(r)$ and $\sinh(r)$, which characterize the squeezing transformation, are given by:
\begin{equation}
    \cosh(r) = \frac{g_1}{\sqrt{g_1^2 - g_2^2}}, \quad \sinh(r) = \frac{g_2}{\sqrt{g_1^2 - g_2^2}}.
\end{equation}
These results demonstrate how the squeezing transformation not only removes the diamagnetic term but also establishes a direct connection between the physical parameters of the system and the mathematical structure of the Jaynes-Cummings model.

Finally, to establish the complete relationship between the Jaynes-Cummings Hamiltonian and the Hamiltonian of the anisotropic quantum Rabi model with the diamagnetic term, Eq.~(\ref{dia}), for the case $g_2 < g_1$, it is essential to recall the relation $\hat{H}_{\text{JCM}} = \hat{S}^\dagger(r) \hat{H} \hat{S}(r)$. This relation is fundamental for finding the eigenvalues of $\hat{H}$ in this parameter regime. Therefore, by multiplying the relation $\hat{S}^\dagger(r) \hat{H} \hat{S}(r)$ by $\hat{S}(r)$, we obtain:
\begin{equation}
\hat{H} \hat{S}(r) \ket{\psi_n^{\text{JCM}}, \pm} = E_{n, \pm}^{\text{JCM}} \hat{S}(r) \ket{\psi_n^{\text{JCM}}, \pm},
\end{equation}
where the eigenvalues $E_{n, \pm}^{\text{JCM}}$ and the eigenvectors $\ket{\psi_n^{\text{JCM}}, \pm}$ are those of the Jaynes-Cummings Hamiltonian, determined by:
\begin{equation}
    \begin{split}
    E_{n, \pm}^{\text{JCM}} &= \omega_{\text{eff}} \left( n + \frac{1}{2} \right) \pm \frac{1}{2} \sqrt{ (\omega_0 - \omega_{\text{eff}})^2 + 4 g_{\text{eff}}^{2} (n+1) } + f_0, \\
    \hat{S}(r) \ket{\psi_{n}^{\text{JCM}}, +} &= \hat{S}(r) \left( \cos(\theta_n) \ket{n, \uparrow} + \sin(\theta_n) \ket{n+1, \downarrow} \right), \\
    \hat{S}(r) \ket{\psi_{n}^{\text{JCM}}, -} &= \hat{S}(r) \left( -\sin(\theta_n) \ket{n, \uparrow} + \cos(\theta_n) \ket{n+1, \downarrow} \right),  
    \end{split}
\end{equation}
with $\tan(2\theta_n) = \frac{2 g_{\text{eff}} \sqrt{n+1}}{\omega_0 - \omega_{\text{eff}}}$. Here, $\ket{n, \uparrow} = \ket{n} \otimes \ket{\uparrow}$ and $\ket{n+1, \downarrow} = \ket{n+1} \otimes \ket{\downarrow}$ are the basis vectors in the Fock space and the atomic subspace, respectively. Clearly, the eigenvalues of $\hat{H}$, given by Eq.~(\ref{dia}), are the same as those of the Jaynes-Cummings Hamiltonian. Moreover, the eigenstates of $\hat{H}$ are connected to those of the Jaynes-Cummings model through the action of the squeeze operator $\hat{S}(r)$ on the eigenstates $\ket{\psi_n^{\text{JCM}},\pm}$.

%------------------------------------
\subsection{Anti-Jaynes-Cummings Model}
On the other hand, starting from Eq.~(\ref{H_S}), if the condition $\frac{\mu}{\nu} = \frac{g_2}{g_1}$ is imposed, where $\mu = \cosh(r)$ and $\nu = \sinh(r)$ are the hyperbolic functions associated with the squeezing parameter $r$, and considering the parameter region defined by $g_1 < g_2$, the Hamiltonian $\hat{H}_S$ can be expressed as:
\begin{equation}
    \hat{H}_{\text{AJCM}} = \tilde{\omega}_\text{eff} \hat{a}^\dagger \hat{a} + \frac{\omega_{0}}{2} \hat{\sigma}_z + \tilde{g}_{\text{eff}} \left( \hat{a} \hat{\sigma}_- + \hat{a}^\dagger \hat{\sigma}_+ \right) + \tilde{f}_0,
\end{equation}
where
\begin{equation}
    \tilde{\omega}_{\text{eff}} = \omega_D \left( \frac{g_2^2 - g_1^2}{g_1^2 + g_2^2} \right), \quad \tilde{g}_{\text{eff}} = \sqrt{g_2^2 - g_1^2}, \quad \text{and} \quad \tilde{f}_0 = \omega_D \left( \frac{g_1 g_2 - g_1^2}{g_1^2 + g_2^2} \right).
\end{equation}
This Hamiltonian can be identified as the anti-Jaynes-Cummings Hamiltonian. In this parameter region, where $g_1 < g_2$, Eq.~(\ref{condicion}) remains unchanged, while the hyperbolic functions $\cosh(r)$ and $\sinh(r)$ are given by:
\begin{equation}
    \cosh(r) = \frac{g_2}{\sqrt{g_2^2 - g_1^2}}, \quad \sinh(r) = \frac{g_1}{\sqrt{g_2^2 - g_1^2}}.
\end{equation}
Finally, to establish the complete relationship between the anti-Jaynes-Cummings Hamiltonian and the Hamiltonian of the anisotropic quantum Rabi model with the diamagnetic term, Eq.~(\ref{dia}), for the case $g_1 < g_2$, it is essential to recall the relation $\hat{H}_{\text{AJCM}} = \hat{S}^\dagger(r) \hat{H} \hat{S}(r)$.  Therefore, by multiplying the relation $\hat{S}^\dagger(r) \hat{H} \hat{S}(r)$ by $\hat{S}(r)$, we obtain:
\begin{equation}
    \begin{split}
    E_{n, \pm}^{\text{AJCM}} &= \tilde{\omega}_{\text{eff}} \left( n + \frac{1}{2} \right) \pm \frac{1}{2} \sqrt{ (\omega_0 + \tilde{\omega}_{\text{eff}})^2 + 4 \tilde{g}_{\text{eff}}^2 (n+1) } + \tilde{f}_0, \\
    \hat{S}(r) \ket{\psi_{n, +}^{\text{AJCM}}} &= \hat{S}(r) \left( \cos(\theta_n) \ket{n+1, \uparrow} + \sin(\theta_n) \ket{n, \downarrow} \right), \\
    \hat{S}(r) \ket{\psi_{n, -}^{\text{AJCM}}} &= \hat{S}(r) \left( -\sin(\theta_n) \ket{n+1, \uparrow} + \cos(\theta_n) \ket{n, \downarrow} \right),
    \end{split}
\end{equation}
where $\tan(2\theta_n) = \frac{2 \tilde{g}_{\text{eff}} \sqrt{n+1}}{\omega_0 + \tilde{\omega}_{\text{eff}}}$. The eigenvectors are given by linear combinations of the states $\ket{n+1, \uparrow}$ and $\ket{n, \downarrow}$. This structure aligns with the anti-Jaynes-Cummings model, where the anti-resonant coupling links the states $\ket{n+1, \uparrow}$ and $\ket{n, \downarrow}$, in contrast to the Jaynes-Cummings model, which connects $\ket{n, \uparrow}$ and $\ket{n+1, \downarrow}$. Once again, the eigenvalues are identical to those of the anti-Jaynes-Cummings Hamiltonian, and the eigenvectors are related via the squeeze operator, $\hat{S}(r)\ket{\psi_n^{\text{AJCM}},\pm}$.

%------------------------------------
\section{Results and Discussion}
In this section, we analyze the eigenvalues and atomic inversion for the anisotropic quantum Rabi model with diamagnetic term in the two distinct parameter regimes: (a) $g_2 < g_1$, corresponding to the Jaynes-Cummings model, and (b) $g_1 < g_2$, associated with the anti-Jaynes-Cummings model. We first discuss the eigenvalues in both regimes and then examine the behavior of the atomic inversion.

\begin{figure}[h]
    \begin{center}
    \includegraphics{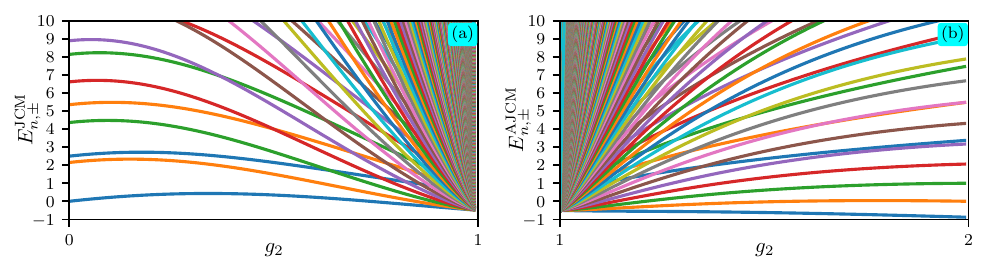}
    \end{center}
    \caption{Energy levels of the anisotropic quantum Rabi model for the first ten states ($n, \pm = 10$) are plotted as a function of the coupling parameter $g_2$, with fixed parameters $\omega_D = 2.5$, $\omega_0 = 1.0$, and $g_1 = 1.0$. In Panel (a), the eigenvalues correspond to the Jaynes-Cummings regime ($g_2 < g_1$), where the energy levels $E_{n, \pm}^{\text{JCM}}$ are derived from Eq.~(\ref{Casos}). Panel (b) illustrates the eigenvalues for the anti-Jaynes-Cummings regime ($g_2 > 1$), with the energy levels $E_{n, \pm}^{\text{AJCM}}$ also determined by Eq.~(\ref{Casos}). This separation highlights the distinct behaviors of the system in the two coupling regimes.}\label{fig_1}
\end{figure}

The eigenvalues $E_{n,\pm}$ of the Hamiltonian $\hat{H}$ are determined by the effective parameters obtained from the squeezing transformation $\hat{S}(r)$ in each parameter regime: (a)~$g_2 < g_1$ and (b)~$g_1 < g_2$ (with $g_1 \neq g_2$). As established in the previous section, these eigenvalues are expressed as:
\begin{equation}\label{Casos}
    E_{n, \pm} = 
    \begin{cases}
        \omega_{\text{eff}} \left( n + \frac{1}{2} \right) \pm \frac{1}{2} \sqrt{ (\omega_0 - \omega_{\text{eff}})^2 + 4 g_{\text{eff}}^{2} (n+1) } + f_0,&\quad\mbox{if}\quad g_2 < g_1,\\
        \tilde{\omega}_{\text{eff}} \left( n + \frac{1}{2} \right) \pm \frac{1}{2} \sqrt{ (\omega_0 + \tilde{\omega}_{\text{eff}})^2 + 4 \tilde{g}_{\text{eff}}^2 (n+1) } + \tilde{f}_0,&\quad\mbox{if}\quad g_1 < g_2,
    \end{cases}
\end{equation}
Figure~\ref{fig_1} shows the first energy levels ($n= 10$) for both the Jaynes-Cummings and anti-Jaynes-Cummings models as a function of the coupling parameter $g_2$, with $g_1 = 1$ fixed. In the regime $g_2 < g_1$, corresponding to the Jaynes-Cummings model, the energy levels $E_{n, \pm}^{\text{JCM}}$ depend on $g_2$, with their structure determined by the effective coupling $g_{\text{eff}} = \sqrt{g_1^2 - g_2^2}$. As $g_2$ approaches $g_1=1$, the energy levels gradually converge, reflecting the diminishing influence of the effective coupling. For $g_1 > g_2$, the system transitions to the anti-Jaynes-Cummings regime, where the energy levels $E_{n, \pm}^{\text{AJCM}}$ are governed by the effective coupling $\tilde{g}_{\text{eff}} = \sqrt{g_2^2 - g_1^2}$, resulting in a modified level structure due to the interaction. The transition between the two regimes is most evident at $g_2\rightarrow g_1$, where the effective coupling changes sign, highlighting the role of anisotropy in the quantum Rabi model and its connection to the two distinct regimes. Although the Fig.~\ref{fig_1} focuses on the first ten energy levels, it is notable that higher-order eigenvalues exhibit a clear trend: they either approach $g_1$ or emerge from it as $g_2$ varies, illustrating the influence of the coupling parameters on the energy spectrum and providing insight into the system's response to changes in $g_2$.

To conclude this section, we present an analysis of the atomic inversion for the anisotropic quantum Rabi model with diamagnetic term in the two distinct coupling regimes. The atomic inversion, denoted as $W(t)$, is a fundamental quantity that characterizes the dynamics of the system. It is defined as the difference in population between the atomic states $\ket{\uparrow}$ and $\ket{\downarrow}$, and is mathematically expressed as: $W(t) = \braket{\hat{\sigma}_z(t)}$. This quantity provides insight into the temporal evolution of the atomic populations and serves as a key indicator of the system's behavior under different coupling conditions. In the two parameter regimes under consideration, the atomic inversion is given by: $W(t) = \braket{\psi(0)|\hat{U}^\dagger(t)\hat{\sigma}_z\hat{U}(t)|\psi(0)}$, where $\hat{U}(t)$ represents the time evolution operator. This operator is defined as:
$\hat{U}(t) = \exp[-i\hat{H}t]$. Here, $\ket{\psi(0)}$ denotes the initial state of the system, which is the tensor product of the field state and the initial atomic state. The atomic inversion captures the interplay between the atomic and field degrees of freedom, reflecting the influence of the coupling parameters $g_1$ and $g_2$. For each parameter regime, the atomic inversion takes the following form:
\begin{equation}
W(t) = 
\begin{cases}
\braket{\psi(0)|\hat{S}(r)\hat{U}_{\text{JCM}}^\dagger(t)\hat{S}(r)\,\hat{\sigma}_z\,\hat{S}(r)\hat{U}_{\text{JCM}}(t)\hat{S}^\dagger(r)|\psi(0)}, & \text{if } g_2 < g_1, \\
\braket{\psi(0)|\hat{S}(r)\hat{U}_{\text{AJCM}}^\dagger(t)\hat{S}(r)\,\hat{\sigma}_z\,\hat{S}(r)\hat{U}_{\text{AJCM}}(t)\hat{S}^\dagger(r)|\psi(0)}, & \text{if } g_1 < g_2.
\end{cases}
\end{equation}
The time evolution operator for the anisotropic quantum Rabi model with diamagnetic term is expressed as:
\begin{equation}
\hat{U}(t)=
\begin{cases}
\hat{S}(r)\hat{U}_{\text{JCM}}(t)\hat{S}^\dagger(r), &\quad \text{for } g_2 < g_1,\\
\hat{S}(r)\hat{U}_{\text{AJCM}}(t)\hat{S}^\dagger(r), &\quad \text{for } g_1 < g_2.
\end{cases}
\end{equation}
The evolution operators corresponding to the Jaynes-Cummings and anti-Jaynes-Cummings Hamiltonians are given by:
\begin{equation}
\begin{split}
\hat{U}_{\text{JCM}} &= \exp\left[-it\left(\hat{a}^\dagger\hat{a}+\frac{\hat{\sigma}_z}{2}\right)\right]
\begin{bmatrix}
\hat{U}_{11}^{\text{JCM}}(t) & \hat{U}_{12}^{\text{JCM}}(t) \\
\hat{U}_{21}^{\text{JCM}}(t) & \hat{U}_{22}^{\text{JCM}}(t)
\end{bmatrix}, \\
\hat{U}_{\text{AJCM}} &= \exp\left[-it\left(\hat{a}^\dagger\hat{a}-\frac{\hat{\sigma}_z}{2}\right)\right]
\begin{bmatrix}
\hat{U}_{11}^{\text{AJCM}}(t) & \hat{U}_{12}^{\text{AJCM}}(t) \\
\hat{U}_{21}^{\text{AJCM}}(t) & \hat{U}_{22}^{\text{AJCM}}(t)
\end{bmatrix}.
\end{split}
\end{equation}
The matrix elements of the evolution operators are explicitly given by:
\begin{equation}
\begin{split}
\hat{U}_{11}^{\text{JCM}}(t) &= \cos\left(\frac{\Omega_{\hat{n}+1}t}{2}\right) - i\frac{\left(\omega_0 - \omega_{\text{eff}}\right)}{\Omega_{\hat{n}+1}}\sin\left(\frac{\Omega_{\hat{n}+1}t}{2}\right), \\
\hat{U}_{12}^{\text{JCM}}(t) &= -ig_{\text{eff}}\frac{\sin\left(\frac{\Omega_{\hat{n}+1}t}{2}\right)}{\Omega_{\hat{n}+1}}\hat{a}, \\
\hat{U}_{21}^{\text{JCM}}(t) &= -ig_{\text{eff}}\hat{a}^\dagger\frac{\sin\left(\frac{\Omega_{\hat{n}+1}t}{2}\right)}{\Omega_{\hat{n}+1}}, \\
\hat{U}_{22}^{\text{JCM}}(t) &= \cos\left(\frac{\Omega_{\hat{n}}t}{2}\right) + i\frac{\left(\omega_0 - \omega_{\text{eff}}\right)}{\Omega_{\hat{n}}}\sin\left(\frac{\Omega_{\hat{n}}t}{2}\right),
\end{split}
\end{equation}
where $\Omega_{\hat{n}} = \sqrt{(\omega_0 - \omega_{\text{eff}})^2 + 4g_{\text{eff}}^2\hat{n}}$ (with $\hat{n} = \hat{a}^\dagger\hat{a}$). Similarly, for the anti-Jaynes-Cummings model:
\begin{equation}
\begin{split}
\hat{U}_{11}^{\text{AJCM}}(t) &= \cos\left(\frac{\tilde{\Omega}_{\hat{n}}t}{2}\right) - i\frac{\left(\omega_0 + \tilde{\omega}_{\text{eff}}\right)}{\tilde{\Omega}_{\hat{n}+1}}\sin\left(\frac{\tilde{\Omega}_{\hat{n}+1}t}{2}\right), \\
\hat{U}_{12}^{\text{AJCM}}(t) &= -i\tilde{g}_{\text{eff}}\hat{a}^\dagger\frac{\sin\left(\frac{\tilde{\Omega}_{\hat{n}+1}t}{2}\right)}{\tilde{\Omega}_{\hat{n}+1}}, \\
\hat{U}_{21}^{\text{AJCM}}(t) &= -i\tilde{g}_{\text{eff}}\frac{\sin\left(\frac{\tilde{\Omega}_{\hat{n}+1}t}{2}\right)}{\tilde{\Omega}_{\hat{n}+1}}\hat{a}, \\
\hat{U}_{22}^{\text{AJCM}}(t) &= \cos\left(\frac{\tilde{\Omega}_{\hat{n}+1}t}{2}\right) + i\frac{\left(\omega_0 + \tilde{\omega}_{\text{eff}}\right)}{\tilde{\Omega}_{\hat{n}+1}}\sin\left(\frac{\tilde{\Omega}_{\hat{n}+1}t}{2}\right),
\end{split}
\end{equation}
with $\tilde{\Omega}_{\hat{n}} = \sqrt{(\omega_0 + \tilde{\omega}_{\text{eff}})^2 + 4\tilde{g}_{\text{eff}}^2\hat{n}}$. 
\begin{figure}[h!]
    \centering
    \includegraphics{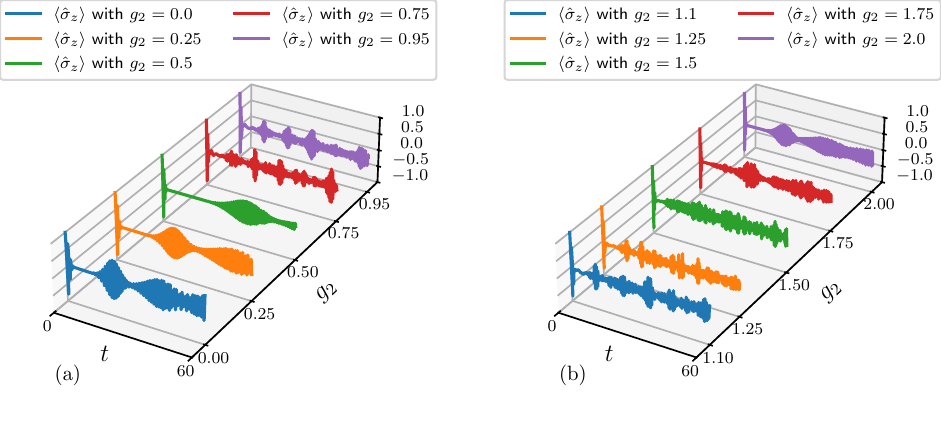}
    \caption{Dynamics of the atomic inversion in the anisotropic quantum Rabi model. The figure illustrates the temporal evolution of the atomic inversion $W(t)$ for the two distinct coupling regimes: (a)~$g_2 < g_1$ and (b)~$g_1 < g_2$. The initial state of the system is chosen as $\ket{\psi(0)} = \ket{\alpha} \otimes \ket{\uparrow} = \ket{\alpha, \uparrow}$, where $\ket{\alpha}$ represents a coherent state of the field with $\alpha = 3 + i$.}
    \label{fig_2}
\end{figure}

Figure~\ref{fig_2} illustrates the time evolution of the atomic inversion $W(t)$ for the anisotropic quantum Rabi model, including the diamagnetic term, in two distinct coupling regimes: (a) $g_2 < g_1$ and (b) $g_1 < g_2$. The initial state of the system is chosen as $\ket{\psi(0)} = \ket{\alpha} \otimes \ket{\uparrow} = \ket{\alpha, \uparrow}$, where $\ket{\alpha}$ represents a coherent state of the field with $\alpha = 3 + i$. As can be observed, the typical revivals of the Jaynes-Cummings model are evident and depend on the effective coupling constant $g_{\text{eff}} = \sqrt{g_1^2 - g_2^2}$. These revivals are plotted for the values $g_2 = [0.0, 0.25, 0.5, 0.75, 0.95]$, and the oscillations reflect the coherent exchange of energy between the atomic and field components. In contrast, in regime (b), we observe that these revivals begin to stabilize for values of $g_2$ significantly larger than $g_1$, corresponding to the anti-Jaynes-Cummings model. In this case, the revivals occur more rapidly and stabilize at larger values of $g_2$, characterized by the effective coupling $\tilde{g}_{\text{eff}} = \sqrt{g_2^2 - g_1^2}$. This change in behavior highlights the sensitivity of the system to the relative strengths of the coupling parameters $g_1$ and $g_2$. For regime (b), we consider the cases $g_2 = [1.1, 1.25, 1.5, 1.75, 2.0]$. Overall, the results demonstrate how the anisotropic quantum Rabi model with diamagnetic term exhibits distinct dynamical behaviors depending on the relative magnitudes of the effective coupling parameters $g_{\text{eff}}$ and $\tilde{g}_{\text{eff}}$.

%------------------------------------
\section{Conclusions}
It has been demonstrated that, by judiciously tuning the diamagnetic amplitude, the anisotropic quantum Rabi model can be reduced to either the Jaynes-Cummings model or the anti-Jaynes-Cummings model through the application of a squeezing transformation. Specifically, when the condition $\frac{\mu}{\nu} = \frac{g_1}{g_2}$ or $\frac{\mu}{\nu} = \frac{g_2}{g_1}$ is satisfied, where $\mu = \cosh(r)$ and $\nu = \sinh(r)$ are the hyperbolic functions associated with the squeezing parameter $r$, the Hamiltonian of the anisotropic quantum Rabi model transforms into the Jaynes-Cummings Hamiltonian for $g_2 < g_1$ or the anti-Jaynes-Cummings Hamiltonian for $g_1 < g_2$. In the case of the standard quantum Rabi model (i.e., when $g_1 = g_2$), the system cannot be reduced to either the Jaynes-Cummings or anti-Jaynes-Cummings models. However, the squeezing transformation still allows us to eliminate the diamagnetic term, thereby removing the $A^2$ interaction from the Hamiltonian. This result highlights the versatility of the squeezing transformation in simplifying the anisotropic quantum Rabi model and its connection to well-known models in quantum optics.

\bibliography{test}

%apsrev4-2.bst 2019-01-14 (MD) hand-edited version of apsrev4-1.bst
%Control: key (0)
%Control: author (8) initials jnrlst
%Control: editor formatted (1) identically to author
%Control: production of article title (0) allowed
%Control: page (0) single
%Control: year (1) truncated
%Control: production of eprint (0) enabled
\begin{thebibliography}{22}%
\makeatletter
\providecommand \@ifxundefined [1]{%
 \@ifx{#1\undefined}
}%
\providecommand \@ifnum [1]{%
 \ifnum #1\expandafter \@firstoftwo
 \else \expandafter \@secondoftwo
 \fi
}%
\providecommand \@ifx [1]{%
 \ifx #1\expandafter \@firstoftwo
 \else \expandafter \@secondoftwo
 \fi
}%
\providecommand \natexlab [1]{#1}%
\providecommand \enquote  [1]{``#1''}%
\providecommand \bibnamefont  [1]{#1}%
\providecommand \bibfnamefont [1]{#1}%
\providecommand \citenamefont [1]{#1}%
\providecommand \href@noop [0]{\@secondoftwo}%
\providecommand \href [0]{\begingroup \@sanitize@url \@href}%
\providecommand \@href[1]{\@@startlink{#1}\@@href}%
\providecommand \@@href[1]{\endgroup#1\@@endlink}%
\providecommand \@sanitize@url [0]{\catcode `\\12\catcode `\$12\catcode `\&12\catcode `\#12\catcode `\^12\catcode `\_12\catcode `\%12\relax}%
\providecommand \@@startlink[1]{}%
\providecommand \@@endlink[0]{}%
\providecommand \url  [0]{\begingroup\@sanitize@url \@url }%
\providecommand \@url [1]{\endgroup\@href {#1}{\urlprefix }}%
\providecommand \urlprefix  [0]{URL }%
\providecommand \Eprint [0]{\href }%
\providecommand \doibase [0]{https://doi.org/}%
\providecommand \selectlanguage [0]{\@gobble}%
\providecommand \bibinfo  [0]{\@secondoftwo}%
\providecommand \bibfield  [0]{\@secondoftwo}%
\providecommand \translation [1]{[#1]}%
\providecommand \BibitemOpen [0]{}%
\providecommand \bibitemStop [0]{}%
\providecommand \bibitemNoStop [0]{.\EOS\space}%
\providecommand \EOS [0]{\spacefactor3000\relax}%
\providecommand \BibitemShut  [1]{\csname bibitem#1\endcsname}%
\let\auto@bib@innerbib\@empty
%</preamble>
\bibitem [{\citenamefont {{J}aynes}\ and\ \citenamefont {{C}ummings}(1963)}]{JC_1963}%
  \BibitemOpen
  \bibfield  {author} {\bibinfo {author} {\bibfnamefont {E.~T.}\ \bibnamefont {{J}aynes}}\ and\ \bibinfo {author} {\bibfnamefont {F.~W.}\ \bibnamefont {{C}ummings}},\ }\bibfield  {title} {\bibinfo {title} {Comparison of quantum and semiclassical radiation theories with application to the beam maser},\ }\href {https://doi.org/10.1109/PROC.1963.1664} {\bibfield  {journal} {\bibinfo  {journal} {Proc.~IEEE}\ }\textbf {\bibinfo {volume} {51}},\ \bibinfo {pages} {89} (\bibinfo {year} {1963})}\BibitemShut {NoStop}%
\bibitem [{\citenamefont {Gerry}\ and\ \citenamefont {Knight}(2004)}]{Gerry_Book}%
  \BibitemOpen
  \bibfield  {author} {\bibinfo {author} {\bibfnamefont {C.~C.}\ \bibnamefont {Gerry}}\ and\ \bibinfo {author} {\bibfnamefont {P.~L.}\ \bibnamefont {Knight}},\ }\href {https://doi.org/10.1017/CBO9780511791239} {\emph {\bibinfo {title} {Introductory Quantum Optics}}}\ (\bibinfo  {publisher} {Cambridge University Press},\ \bibinfo {year} {2004})\BibitemShut {NoStop}%
\bibitem [{\citenamefont {Larson}\ and\ \citenamefont {Mavrogordatos}(2022)}]{Larson_2022}%
  \BibitemOpen
  \bibfield  {author} {\bibinfo {author} {\bibfnamefont {J.}~\bibnamefont {Larson}}\ and\ \bibinfo {author} {\bibfnamefont {T.~K.}\ \bibnamefont {Mavrogordatos}},\ }\href {https://api.semanticscholar.org/CorpusID:245226335} {\emph {\bibinfo {title} {The {J}aynes–{C}ummings Model and Its Descendants}}}\ (\bibinfo  {publisher} {Institute of Physics Publishing},\ \bibinfo {year} {2022})\BibitemShut {NoStop}%
\bibitem [{\citenamefont {Meher}\ and\ \citenamefont {Sivakumar}(2022)}]{Nilakanta}%
  \BibitemOpen
  \bibfield  {author} {\bibinfo {author} {\bibfnamefont {N.}~\bibnamefont {Meher}}\ and\ \bibinfo {author} {\bibfnamefont {S.}~\bibnamefont {Sivakumar}},\ }\bibfield  {title} {\bibinfo {title} {A review on quantum information processing in cavities},\ }\href {https://doi.org/10.1140/epjp/s13360-022-03172-x} {\bibfield  {journal} {\bibinfo  {journal} {Eur. Phys. J. Plus}\ }\textbf {\bibinfo {volume} {137}},\ \bibinfo {pages} {985} (\bibinfo {year} {2022})}\BibitemShut {NoStop}%
\bibitem [{\citenamefont {Meher}\ and\ \citenamefont {Sivakumar}(2018)}]{Nilakanta2}%
  \BibitemOpen
  \bibfield  {author} {\bibinfo {author} {\bibfnamefont {N.}~\bibnamefont {Meher}}\ and\ \bibinfo {author} {\bibfnamefont {S.}~\bibnamefont {Sivakumar}},\ }\bibfield  {title} {\bibinfo {title} {Number state filtered coherent states},\ }\href {https://doi.org/10.1007/s11128-018-1995-6} {\bibfield  {journal} {\bibinfo  {journal} {Quantum Inf. Process.}\ }\textbf {\bibinfo {volume} {17}},\ \bibinfo {pages} {233} (\bibinfo {year} {2018})}\BibitemShut {NoStop}%
\bibitem [{\citenamefont {Meher}\ \emph {et~al.}(2017)\citenamefont {Meher}, \citenamefont {Sivakumar},\ and\ \citenamefont {Panigrahi}}]{Nilakanta3}%
  \BibitemOpen
  \bibfield  {author} {\bibinfo {author} {\bibfnamefont {N.}~\bibnamefont {Meher}}, \bibinfo {author} {\bibfnamefont {S.}~\bibnamefont {Sivakumar}},\ and\ \bibinfo {author} {\bibfnamefont {P.~K.}\ \bibnamefont {Panigrahi}},\ }\bibfield  {title} {\bibinfo {title} {Duality and quantum state engineering in cavity arrays},\ }\href {https://doi.org/10.1038/s41598-017-08569-8} {\bibfield  {journal} {\bibinfo  {journal} {Sci. Rep.}\ }\textbf {\bibinfo {volume} {7}},\ \bibinfo {pages} {9251} (\bibinfo {year} {2017})}\BibitemShut {NoStop}%
\bibitem [{\citenamefont {Crisp}(1991)}]{Crisp}%
  \BibitemOpen
  \bibfield  {author} {\bibinfo {author} {\bibfnamefont {M.~D.}\ \bibnamefont {Crisp}},\ }\bibfield  {title} {\bibinfo {title} {Interaction of a charged harmonic oscillator with a single quantized electromagnetic field mode},\ }\href {https://doi.org/10.1103/PhysRevA.44.563} {\bibfield  {journal} {\bibinfo  {journal} {Phys. Rev. A}\ }\textbf {\bibinfo {volume} {44}},\ \bibinfo {pages} {563} (\bibinfo {year} {1991})}\BibitemShut {NoStop}%
\bibitem [{\citenamefont {Crisp}(1993)}]{Crisp_1993}%
  \BibitemOpen
  \bibfield  {author} {\bibinfo {author} {\bibfnamefont {M.~D.}\ \bibnamefont {Crisp}},\ }\bibinfo {title} {Ed {J}aynes’ steak dinner problem ii},\ in\ \href {https://doi.org/10.1017/CBO9780511524448} {\emph {\bibinfo {booktitle} {Physics and Probability: Essays in Honor of Edwin T. {J}aynes}}},\ \bibinfo {editor} {edited by\ \bibinfo {editor} {\bibfnamefont {W.~T.}\ \bibnamefont {Grandy}, \bibfnamefont {Jr}}\ and\ \bibinfo {editor} {\bibfnamefont {P.~W.}\ \bibnamefont {Milonni}}}\ (\bibinfo  {publisher} {Cambridge University Press},\ \bibinfo {year} {1993})\ p.\ \bibinfo {pages} {81–90}\BibitemShut {NoStop}%
\bibitem [{\citenamefont {Kockum}\ \emph {et~al.}(2019)\citenamefont {Kockum}, \citenamefont {Miranowicz}, \citenamefont {Liberato}, \citenamefont {Savasta},\ and\ \citenamefont {Nori}}]{Kockum_2019b}%
  \BibitemOpen
  \bibfield  {author} {\bibinfo {author} {\bibfnamefont {A.~F.}\ \bibnamefont {Kockum}}, \bibinfo {author} {\bibfnamefont {A.}~\bibnamefont {Miranowicz}}, \bibinfo {author} {\bibfnamefont {S.~D.}\ \bibnamefont {Liberato}}, \bibinfo {author} {\bibfnamefont {S.}~\bibnamefont {Savasta}},\ and\ \bibinfo {author} {\bibfnamefont {F.}~\bibnamefont {Nori}},\ }\bibfield  {title} {\bibinfo {title} {Ultrastrong coupling between light and matter},\ }\href {https://doi.org/10.1038/s42254-018-0006-2} {\bibfield  {journal} {\bibinfo  {journal} {Nat. Rev. Phys.}\ }\textbf {\bibinfo {volume} {1}},\ \bibinfo {pages} {19} (\bibinfo {year} {2019})}\BibitemShut {NoStop}%
\bibitem [{\citenamefont {Salado-Mejía}\ \emph {et~al.}(2021)\citenamefont {Salado-Mejía}, \citenamefont {Román-Ancheyta}, \citenamefont {Soto-Eguibar},\ and\ \citenamefont {Moya-Cessa}}]{Salado_2021}%
  \BibitemOpen
  \bibfield  {author} {\bibinfo {author} {\bibfnamefont {M.}~\bibnamefont {Salado-Mejía}}, \bibinfo {author} {\bibfnamefont {R.}~\bibnamefont {Román-Ancheyta}}, \bibinfo {author} {\bibfnamefont {F.}~\bibnamefont {Soto-Eguibar}},\ and\ \bibinfo {author} {\bibfnamefont {H.~M.}\ \bibnamefont {Moya-Cessa}},\ }\bibfield  {title} {\bibinfo {title} {Spectroscopy and critical quantum thermometry in the ultrastrong coupling regime},\ }\href {https://doi.org/10.1088/2058-9565/abdca5} {\bibfield  {journal} {\bibinfo  {journal} {Quantum Science and Technology}\ }\textbf {\bibinfo {volume} {6}},\ \bibinfo {pages} {025010} (\bibinfo {year} {2021})}\BibitemShut {NoStop}%
\bibitem [{\citenamefont {Moya-cessa}\ \emph {et~al.}(1999)\citenamefont {Moya-cessa}, \citenamefont {Roversi}, \citenamefont {Dutra},\ and\ \citenamefont {Vidiella-barranco}}]{MoyaCessaPRA}%
  \BibitemOpen
  \bibfield  {author} {\bibinfo {author} {\bibfnamefont {H.}~\bibnamefont {Moya-cessa}}, \bibinfo {author} {\bibfnamefont {J.~A.}\ \bibnamefont {Roversi}}, \bibinfo {author} {\bibfnamefont {S.~M.}\ \bibnamefont {Dutra}},\ and\ \bibinfo {author} {\bibfnamefont {A.}~\bibnamefont {Vidiella-barranco}},\ }\bibfield  {title} {\bibinfo {title} {Recovering coherence from decoherence: A method of quantum-state reconstruction},\ }\href {https://doi.org/10.1103/PhysRevA.60.4029} {\bibfield  {journal} {\bibinfo  {journal} {Phys. Rev. A}\ }\textbf {\bibinfo {volume} {60}},\ \bibinfo {pages} {4029} (\bibinfo {year} {1999})}\BibitemShut {NoStop}%
\bibitem [{\citenamefont {Braak}(2011)}]{Braak}%
  \BibitemOpen
  \bibfield  {author} {\bibinfo {author} {\bibfnamefont {D.}~\bibnamefont {Braak}},\ }\bibfield  {title} {\bibinfo {title} {Integrability of the {R}abi model},\ }\href {https://doi.org/10.1103/PhysRevLett.107.100401} {\bibfield  {journal} {\bibinfo  {journal} {Phys. Rev. Lett.}\ }\textbf {\bibinfo {volume} {107}},\ \bibinfo {pages} {100401} (\bibinfo {year} {2011})}\BibitemShut {NoStop}%
\bibitem [{\citenamefont {Swain}(1973)}]{Swain}%
  \BibitemOpen
  \bibfield  {author} {\bibinfo {author} {\bibfnamefont {S.}~\bibnamefont {Swain}},\ }\bibfield  {title} {\bibinfo {title} {Continued fraction expressions for the eigensolutions of the hamiltonian describing the interaction between a single atom and a single field mode: comparisons with the rotating wave solutions},\ }\href {https://doi.org/10.1088/0305-4470/6/12/016} {\bibfield  {journal} {\bibinfo  {journal} {J. Phys. A}\ }\textbf {\bibinfo {volume} {6}},\ \bibinfo {pages} {1919} (\bibinfo {year} {1973})}\BibitemShut {NoStop}%
\bibitem [{\citenamefont {Moya-Cessa}\ \emph {et~al.}(2000)\citenamefont {Moya-Cessa}, \citenamefont {Vidiella-Barranco}, \citenamefont {Roversi},\ and\ \citenamefont {Dutra}}]{Moya_2000}%
  \BibitemOpen
  \bibfield  {author} {\bibinfo {author} {\bibfnamefont {H.}~\bibnamefont {Moya-Cessa}}, \bibinfo {author} {\bibfnamefont {A.}~\bibnamefont {Vidiella-Barranco}}, \bibinfo {author} {\bibfnamefont {J.}~\bibnamefont {Roversi}},\ and\ \bibinfo {author} {\bibfnamefont {S.}~\bibnamefont {Dutra}},\ }\bibfield  {title} {\bibinfo {title} {Unitary transformation approach for the trapped ion dynamics},\ }\href {https://doi.org/10.1088/1464-4266/2/1/303} {\bibfield  {journal} {\bibinfo  {journal} {J. Opt. B: Quantum Semiclass. Opt.}\ }\textbf {\bibinfo {volume} {2}},\ \bibinfo {pages} {21} (\bibinfo {year} {2000})}\BibitemShut {NoStop}%
\bibitem [{\citenamefont {Chen}\ \emph {et~al.}(2011)\citenamefont {Chen}, \citenamefont {Liu}, \citenamefont {Zhang},\ and\ \citenamefont {Wang}}]{Chen_2011}%
  \BibitemOpen
  \bibfield  {author} {\bibinfo {author} {\bibfnamefont {Q.-H.}\ \bibnamefont {Chen}}, \bibinfo {author} {\bibfnamefont {T.}~\bibnamefont {Liu}}, \bibinfo {author} {\bibfnamefont {Y.-Y.}\ \bibnamefont {Zhang}},\ and\ \bibinfo {author} {\bibfnamefont {K.-L.}\ \bibnamefont {Wang}},\ }\bibfield  {title} {\bibinfo {title} {Exact solutions to the {J}aynes-{C}ummings model without the rotating-wave approximation},\ }\href {https://doi.org/10.1209/0295-5075/96/14003} {\bibfield  {journal} {\bibinfo  {journal} {EPL}\ }\textbf {\bibinfo {volume} {96}},\ \bibinfo {pages} {14003} (\bibinfo {year} {2011})}\BibitemShut {NoStop}%
\bibitem [{\citenamefont {Xie}\ \emph {et~al.}(2014)\citenamefont {Xie}, \citenamefont {Cui}, \citenamefont {Cao1}, \citenamefont {Amico},\ and\ \citenamefont {Fan}}]{Xie}%
  \BibitemOpen
  \bibfield  {author} {\bibinfo {author} {\bibfnamefont {Q.-T.}\ \bibnamefont {Xie}}, \bibinfo {author} {\bibfnamefont {S.}~\bibnamefont {Cui}}, \bibinfo {author} {\bibfnamefont {J.-P.}\ \bibnamefont {Cao1}}, \bibinfo {author} {\bibfnamefont {L.}~\bibnamefont {Amico}},\ and\ \bibinfo {author} {\bibfnamefont {H.}~\bibnamefont {Fan}},\ }\bibfield  {title} {\bibinfo {title} {Anisotropic {R}abi model},\ }\href {https://doi.org/10.1103/PhysRevX.4.021046} {\bibfield  {journal} {\bibinfo  {journal} {Phys. Rev. X}\ }\textbf {\bibinfo {volume} {4}},\ \bibinfo {pages} {021046} (\bibinfo {year} {2014})}\BibitemShut {NoStop}%
\bibitem [{\citenamefont {Ye}\ \emph {et~al.}(2024)\citenamefont {Ye}, \citenamefont {Wang},\ and\ \citenamefont {Chen}}]{Ye}%
  \BibitemOpen
  \bibfield  {author} {\bibinfo {author} {\bibfnamefont {T.}~\bibnamefont {Ye}}, \bibinfo {author} {\bibfnamefont {C.}~\bibnamefont {Wang}},\ and\ \bibinfo {author} {\bibfnamefont {Q.-H.}\ \bibnamefont {Chen}},\ }\bibfield  {title} {\bibinfo {title} {Anisotropic qubit-photon interactions inducing multiple antibunching-to-bunching transitions of photons},\ }\href {https://doi.org/10.1364/OE.533310} {\bibfield  {journal} {\bibinfo  {journal} {Opt. Express}\ }\textbf {\bibinfo {volume} {32}},\ \bibinfo {pages} {33483} (\bibinfo {year} {2024})}\BibitemShut {NoStop}%
\bibitem [{\citenamefont {Boutakka}\ \emph {et~al.}(2024)\citenamefont {Boutakka}, \citenamefont {Sakhi},\ and\ \citenamefont {Bennai}}]{Boutaka}%
  \BibitemOpen
  \bibfield  {author} {\bibinfo {author} {\bibfnamefont {Z.}~\bibnamefont {Boutakka}}, \bibinfo {author} {\bibfnamefont {Z.}~\bibnamefont {Sakhi}},\ and\ \bibinfo {author} {\bibfnamefont {M.}~\bibnamefont {Bennai}},\ }\bibfield  {title} {\bibinfo {title} {Quantum entanglement in the {R}abi model with the presence of the ${A}^2$ term},\ }\href {https://doi.org/10.1007/s10773-024-05805-6} {\bibfield  {journal} {\bibinfo  {journal} {Int J Theor Phys}\ }\textbf {\bibinfo {volume} {63}},\ \bibinfo {pages} {274} (\bibinfo {year} {2024})}\BibitemShut {NoStop}%
\bibitem [{\citenamefont {Satyanarayana}\ \emph {et~al.}(1989)\citenamefont {Satyanarayana}, \citenamefont {Rice}, \citenamefont {Vyas},\ and\ \citenamefont {Carmichael}}]{Satyanarayana_1989}%
  \BibitemOpen
  \bibfield  {author} {\bibinfo {author} {\bibfnamefont {M.~V.}\ \bibnamefont {Satyanarayana}}, \bibinfo {author} {\bibfnamefont {P.}~\bibnamefont {Rice}}, \bibinfo {author} {\bibfnamefont {R.}~\bibnamefont {Vyas}},\ and\ \bibinfo {author} {\bibfnamefont {H.~J.}\ \bibnamefont {Carmichael}},\ }\bibfield  {title} {\bibinfo {title} {Ringing revivals in the interaction of a two-level atom with squeezed light},\ }\href {https://doi.org/10.1364/JOSAB.6.000228} {\bibfield  {journal} {\bibinfo  {journal} {J. Opt. Soc. Am. B}\ }\textbf {\bibinfo {volume} {6}},\ \bibinfo {pages} {228} (\bibinfo {year} {1989})}\BibitemShut {NoStop}%
\bibitem [{\citenamefont {Rodriguez-Lara}\ \emph {et~al.}(2005)\citenamefont {Rodriguez-Lara}, \citenamefont {Moya-Cessa},\ and\ \citenamefont {Klimov}}]{Anti}%
  \BibitemOpen
  \bibfield  {author} {\bibinfo {author} {\bibfnamefont {B.~M.}\ \bibnamefont {Rodriguez-Lara}}, \bibinfo {author} {\bibfnamefont {H.}~\bibnamefont {Moya-Cessa}},\ and\ \bibinfo {author} {\bibfnamefont {A.~B.}\ \bibnamefont {Klimov}},\ }\bibfield  {title} {\bibinfo {title} {Combining {J}aynes-{C}ummings and anti-{J}aynes-{C}ummings dynamics in a trapped-ion system driven by a laser},\ }\href {https://doi.org/10.1103/PhysRevA.71.023811} {\bibfield  {journal} {\bibinfo  {journal} {Phys. Rev. A}\ }\textbf {\bibinfo {volume} {71}},\ \bibinfo {pages} {023811} (\bibinfo {year} {2005})}\BibitemShut {NoStop}%
\bibitem [{\citenamefont {Bocanegra-Garay}\ \emph {et~al.}(2024)\citenamefont {Bocanegra-Garay}, \citenamefont {Castillo-Celeita}, \citenamefont {Negro}, \citenamefont {Nieto},\ and\ \citenamefont {Gómez-Ruiz}}]{Ivan}%
  \BibitemOpen
  \bibfield  {author} {\bibinfo {author} {\bibfnamefont {I.~A.}\ \bibnamefont {Bocanegra-Garay}}, \bibinfo {author} {\bibfnamefont {M.}~\bibnamefont {Castillo-Celeita}}, \bibinfo {author} {\bibfnamefont {J.}~\bibnamefont {Negro}}, \bibinfo {author} {\bibfnamefont {L.}~\bibnamefont {Nieto}},\ and\ \bibinfo {author} {\bibfnamefont {F.~J.}\ \bibnamefont {Gómez-Ruiz}},\ }\bibfield  {title} {\bibinfo {title} {Exploring supersymmetry: Interchangeability between {J}aynes-{C}ummings and anti-{J}aynes-{C}ummings models},\ }\href {https://doi.org/10.1103/PhysRevResearch.6.043218} {\bibfield  {journal} {\bibinfo  {journal} {Phys. Rev. Research}\ }\textbf {\bibinfo {volume} {6}},\ \bibinfo {pages} {043218} (\bibinfo {year} {2024})}\BibitemShut {NoStop}%
\bibitem [{\citenamefont {Zúñiga-Segundo}\ \emph {et~al.}(2014)\citenamefont {Zúñiga-Segundo}, \citenamefont {Rodríguez-Lara}, \citenamefont {Fernández-C.},\ and\ \citenamefont {Moya-Cessa}}]{David_14}%
  \BibitemOpen
  \bibfield  {author} {\bibinfo {author} {\bibfnamefont {A.}~\bibnamefont {Zúñiga-Segundo}}, \bibinfo {author} {\bibfnamefont {B.~M.}\ \bibnamefont {Rodríguez-Lara}}, \bibinfo {author} {\bibfnamefont {D.~J.}\ \bibnamefont {Fernández-C.}},\ and\ \bibinfo {author} {\bibfnamefont {H.~M.}\ \bibnamefont {Moya-Cessa}},\ }\bibfield  {title} {\bibinfo {title} {Jacobi photonic lattices and their susy partners},\ }\href {https://doi.org/10.1364/OE.22.000987} {\bibfield  {journal} {\bibinfo  {journal} {Optics Express}\ }\textbf {\bibinfo {volume} {22}},\ \bibinfo {pages} {987} (\bibinfo {year} {2014})}\BibitemShut {NoStop}%
\end{thebibliography}%

%%% Make sure to upload the bib file along with the tex file and PDF
%%% Please see the test.bib file for some examples of references

%%% If you don't add the figures in the LaTeX files, please upload them when submitting the article.
%%% Frontiers will add the figures at the end of the provisional pdf automatically
%%% The use of LaTeX coding to draw Diagrams/Figures/Structures should be avoided. They should be external callouts including graphics.

\end{document}